# Correlation between intercalated magnetic layers and superconductivity in pressurized $EuFe_2(As_{0.81}P_{0.19})_2$

Jing Guo[1,2], Qi Wu[2], Ji Feng[1,6], Genfu Chen[2],Tomoko Kagayama[3], Chao Zhang[2], Wei Yi[2], Yanchun Li[4], Xiaodong Li[4], Jing Liu[4], Zheng Jiang[5], Xiangjun Wei[5],Yuying Huang[5], Katsuya Shimizhu[3], Liling Sun[2,6]† & Zhongxian Zhao[2,6]†

[1]*International Center for Quantum Materials, School of Physics, Peking University, Beijing 100871, China*

[2]*Institute of Physics and Beijing National Laboratory for Condensed Matter Physics, Chinese Academy of Sciences, Beijing 100190, China*

[3]*KYOKUGEN, Center for Science and Technology under Extreme Conditions, Osaka University, Machikaneyama-cho, Toyonaka, Osaka 560-8531, Japan*

[4]*Institute of High Energy Physics, Chinese Academy of Sciences, Beijing 100039, China*

[5]*Shanghai Synchrotron Radiation Facilities, Shanghai Institute of Applied Physics, Chinese Academy of Sciences, Shanghai 201204, China*

[6]*Collaborative Innovation Center of Quantum Matter, Beijing, 100190 & 100871, China*

We report comprehensive high pressure studies on correlation between intercalated magnetic layers and superconductivity in $EuFe_2(As_{0.81}P_{0.19})_2$ single crystal through *in-situ* high pressure resistance, specific heat, X-ray diffraction and X-ray absorption measurements. We find that an unconfirmed magnetic order of the intercalated layers coexists with superconductivity in a narrow pressure range 0-0.5GPa, and then it converts to a ferromagnetic (FM) order at pressure above 0.5 GPa, where its superconductivity is absent. The obtained temperature-pressure phase diagram clearly demonstrates that the unconfirmed magnetic order can emerge from the superconducting state. In stark contrast, the superconductivity cannot develop from the FM state that is evolved from the unconfirmed magnetic state. High pressure X-ray absorption (XAS) measurements reveal that the pressure-induced enhancement of Eu's mean valence plays an important role in suppressing the superconductivity and tuning the transition from the unconfirmed magnetic state to a FM state. The unusual interplay among valence state of Eu ions, magnetism and superconductivity under pressure may shed new light on understanding the role of the intercalated magnetic layers in Fe-based superconductors.

A central issue in the studies of superconductivity has been the relationship between magnetism and superconductivity. For conventional superconductors, it is firmly believed that magnetism is hostility to superconductivity, as pointed out by Matthias rule[1]. While, for unconventional superconductors with layered structures constructed by conducting layers with magnetism and the intercalated layers without magnetism, such as cuperates or most of iron pnictide superconductors, the emergence of the superconductivity is closely connected with a suppressed long-ranged antiferromagnetic (AFM) order of the conducting layers. $EuFe_2As_2$(Eu-122) is a peculiar member in Fe-based superconductors, characterized by the presence of dual AFM orders, one of which is from the FeAs layers at 190 K as usual pnictide superconductors, and the other from the intercalated Eu layers forming at 20 K [2,3]. Eu-122 compound had been the only system with the feature of the intercalated layers with magnetism in the high-Tc superconductors, before the discovery of LiFeOHFeSe superconductor recently [4]. For this kind of systems, what is the role of the magnetism from the intercalated layers in developing superconductivity is an intriguing issue.

At ambient pressure, the pristine Eu-122 is not superconducting (NSC), while phosphorous (P) doping can effectively suppress the AFM order of FeAs layers and induce a superconducting transition in the host sample, meanwhile, tune the AFM order of Eu ions to an alternative magnetic order. The type of the magnetic order of Eu ions that coexists with superconductivity is under debate. Some results show that this magnetic order state is FM state [5-9], while the others exhibit that it is an AFM state [10-11] or a canted AFM state plus spin glass [12].Applying external pressure plays the similar role as that of the chemical pressure induced by P doping, driving the undoped Eu-122 to a superconducting state (SC) for pressures ranging from 2.6 to 2.8 GPa and turning the magnetic state of the intercalated layers changes from the AFM to an unconfirmed magnetic state (UM) simultaneously [13-16]. In this study, we are the first to focus on the intercalated Eu layers in $EuFe_2(As_{0.81}P_{0.19})_2$ superconductor, and demonstrate the pressure-induced evolutions of the magnetic order from the intercalated layers, superconductivity, Eu's valence state and lattice parameter in $EuFe_2(As_{0.81}P_{0.19})_2$. Our results reveal that the UM order can emerge from the superconducting state while the superconductivity cannot develop from the FM state. We propose that the UM-FM transition of the intercalated layers and the SC-NSC transition in pressurized $EuFe_2(As_{0.81}P_{0.19})_2$ are associated with pressure-induced changes of the Eu's valence state and lattice parameters.

The single crystals were synthesized by solid reaction methods, as described elsewhere [17]. The actual chemical composition for the host sample is inspected by inductive coupled plasma-atomic emission spectrometer, the resulting composition of which is $EuFe_2(As_{0.81}P_{0.19})_2$. High pressure is created by diamond anvil cell (DAC). Diamond anvils of 600 μm and 300 μm flat tips were used for the low and high pressure measurements, respectively. The non-magnetic rhenium gaskets were used for different experimental runs. The $EuFe_2(As_{0.81}P_{0.19})_2$ single crystal with dimensions around 150×120×20 μm was loaded in the gasket hole and then confined in the hole by squeezing the two anvils. To achieve hydrostatic and quasi-hydrostatic pressure environments for the sample, silicon oil and NaCl powder were employed as pressure transmitting media for high pressure specific heat, *ac* susceptibility and resistance measurements, respectively. The details of high-pressure electrical resistance and *ac*-calorimetric measurements can be found in Ref [18] and Ref [19]. High-pressure angle dispersive X-ray diffraction (XRD) and X-ray absorption spectra (XAS) measurements were carried out at beamline 4W2 at Beijing Synchrotron Radiation Facility (BSRF) and at beamline14W at Shanghai Synchrotron Radiation Facility (SSRF), respectively. In the XRD measurements, a spring steel gasket with a hole about 120 μm in diameter was used over entire pressure range. The XAS measurements were perform with a side incidence, and a beryllium gasket with the same geometry was used in order to obtain sufficient sample signal. Silicon oil was employed as a pressure-transmitting medium to ensure the sample in a hydrostatic pressure environment in the XRD and XAS measurements. Pressure was determined by ruby fluorescence [20].

Figure1 shows the temperature dependence of the electrical resistance of $EuFe_2(As_{0.81}P_{0.19})_2$ single crystal at different pressures. It can be seen that, at ambient pressure, the host sample shows a sharp resistance drop and presents the superconducting transition at the temperature ($Tc$) of 24.5 K. Magnetic susceptibility data measured at ambient pressure clearly exhibit the coexistence of the superconductivity and the magnetic order of Eu ions (inset of Fig.1a). By applying the pressure ~ 0.2 GPa, the $Tc$ is decreased apparently and the magnetic order of the intercalated layers becomes visible concomitantly at a temperature below the $Tc$ (Fig.1a). Here, we define the onset temperature of this magnetic order as $T_m$. Upon increasing pressure to 0.7 GPa, we find that the sample loses its zero resistance and the sharp transition becomes obliterated. Instead, a clear kink is present. Upon further increasing pressure, the resistance kink shifts to higher temperatures. At 8.9 GPa, the onset temperature of the resistance kink is about 48.3 K (Fig.1b). This resistance kink at pressure above 0.5 GPa has been confirmed by experiments to be a signature for the FM transition of the intercalated layers [14]. Here we define the onset temperature of the FM transition as $T_m'$.

Next, high pressure *ac*-calorimetric and resistance measurements are performed. As shown in Fig.2, the measurements on electrical resistance as a function of temperature show two phase transitions at pressure below 0.5 GPa, one of which is related to the superconducting transition and the other to the PM-UM transition of the intercalated layers. It is reproducible for the results that the *Tc* decreases with increasing pressure and vanishes at 0.7 GPa (Fig.2a), and also is consistent with our first run's measurements (Fig.1a). Note that the transition of the PM-UM at $T_m$ is more pronounced at 0.5 GPa, meanwhile, the resistance of the sample is enhanced around $T_m$. High-pressure specific heat measurement is an effective way to identify the transition of magnetic orders. Figure 2b displays the plot of specific heat ($C_{ac}$) versus temperature at different pressures. We find a clear $C_{ac}$ jump for the pressure-free sample, indicating the existence of the magnetic order in the host sample, in good agreement with our magnetic susceptibility measurements (inset of Fig.1a). Upon increasing pressure, the $C_{ac}$ jump shifts steadily to the higher temperature side.

Moreover, the intensity of the $C_{ac}$ jump is found to grow remarkably at 0.7 and 1.0 GPa. Combining all these high-pressure resistance data, we believe that the kink seen on the temperature-resistance curves is associated with the magnetic transition of the intercalated Eu ions.

We summarize our experimental results in Fig.3. It is found that the $Tc$ shows a decrease with an increasing pressure, while the $T_m$ displays an opposite behavior with pressure. These two temperatures ($Tc$ and $T_m$) come cross at ~0.5 GPa. Our results demonstrate a clear process of strong competition between superconductivity and UM order in pressurized $EuFe_2(As_{0.81}P_{0.19})_2$. Further increasing pressure about 0.52 GPa, the UM state turns to a FM state [14] (inset of Fig.3), and the $T_m'$ is found to increase continuously with pressure up to ~10 GPa. The phase diagram clearly demonstrates that a magnetic order can emerge from the superconducting state, while the superconductivity cannot develop from the ferromagnetic state.

To understand the observed correlation between superconductivity and magnetic orders in pressurized $EuFe_2(As_{0.81}P_{0.19})_2$ sample, we performed high-pressure XRD and XAS measurements in diamond anvil cells. No structural phase transition is found over the entire pressure range investigated (Fig.4a), suggesting that the pressure-induced disappearance of superconductivity is solely driven by the competition of electron phases. Nevertheless, a remarkable reduction (compressed by 13.8 % at ~9 GPa) in $c$ direction is observed. The pressure-induced shrinkage of the crystal lattice means a possible valence change of Eu ions because the volume of trivalent Eu with a $4f^6$ electron shell is smaller than that of divalent Eu with a $4f^7$ electron shell. To further study the basic connection among the valence change of Eu ions, superconductivity and magnetic state in $EuFe_2(As_{0.81}P_{0.19})_2$, we performed high-pressure XAS experiments on Eu-$L_3$ absorption edge. As it can be seen in the inset of Fig.4b, the intensity of the main peak associated to the $Eu^{2+}$goes down, whereas the intensity of the satellite peak related to the $Eu^{3+}$ goes up with increasing pressure. The spectra weight transfer indicates the occurrence of the valence change from the $Eu^{2+}$ to $Eu^{3+}$.We estimated the mean valence ( $v$ ) of Eu ions by the equation of $v = 2+I^{Eu3+}/(I^{Eu2+}+I^{Eu3+})$ [21,22], where $I^{Eu2+}$ and $I^{Eu3+}$ are amplitudes of peaks corresponding to $Eu^{2+}$ and $Eu^{3+}$ on XAS spectrum. The estimated $v$ of Eu ions in $EuFe_2(As_{0.81}P_{0.19})_2$ single crystal was plotted in the main figure of Fig. 4b.Apparently, the $v$ is enhanced from 2.32 at ambient pressure up to 2.45 at 9.5 GPa. We also plot the pressure dependences of $T_m$ and $T_m'$ in the same figure and find that the trend of pressure versus $v$, $T_m$ and $T_m'$ are the same, showing monotonically increased with elevating pressure. Considering that P doping or external pressure can turn the magnetic order state of the intercalated layers from a AFM state to a UM state [5-10,14] where the valence of Eu ions changes from $Eu^{2+}$to $Eu^{2.32+}$ at doping level of 0.38 and $Eu^{2+}$ to $Eu^{2.34+}$ at ~10 GPa [14,17,23], we propose that the enhanced mean valence favors AFM-UM-FM phase transitions. The pressure/doping-induced AFM-UM-FM transitions in Eu-122 system are different from the case in LiFeOHFeSe [24], in which its AFM order state of the intercalated layers remains unchanged under chemical pressure due to no such a kind of valence change of Eu ions as presented in the Eu-122 systems. From dynamic point of view for the formations of superconducting and magnetic phases, our results indicate that the pressure-induced enhancement of UM order state suppresses the superconductivity, and the FM order state, which is evolved from the UM order state when $T_m$ is higher than $Tc$, prevents the formation of the superconducting state conclusively. The enhancement of the $v$ is tightly related to the pressure-induced apparent reduction of the volume (Fig.4a).

In conclusion, the pressure-induced evolutions of the magnetic order in the intercalated layers, superconductivity, Eu's valence state and lattice parameter in $EuFe_2(As_{0.81}P_{0.19})_2$ are clearly presented by comprehensive high pressure studies. We find that, below 0.5 GPa, the $Tc$ is decreased with increasing pressure, while the $T_m$ of the UM phase is increased monotonically. At a critical pressure ~0.5 GPa, superconductivity vanishes, and the UM state is switched into an FM state. Our results demonstrate that the UM state can emerge from the superconducting state and coexist with the superconductivity, but the development of superconducting state is prevented from the formation of ferromagnetic state. Pressure-induced enhancement of the mean valence of Eu ions does not help to stabilize superconductivity but favor the UM-FM transition of the intercalated layers in $EuFe_2(As_{0.81}P_{0.19})_2$.

**Acknowledgements**

The work was supported by the NSF of China (Grant Nos 91321207 and 11427805), 973 projects (Grant Nos 2011CBA00100 and 2010CB923000) and the Strategic Priority Research Program (B) of the Chinese Academy of Sciences (Grant No. XDB07020300).

†To whom correspondence should be addressed.

E-mail: llsun@iphy.ac.cn and zhxzhao@iphy.ac.cn .

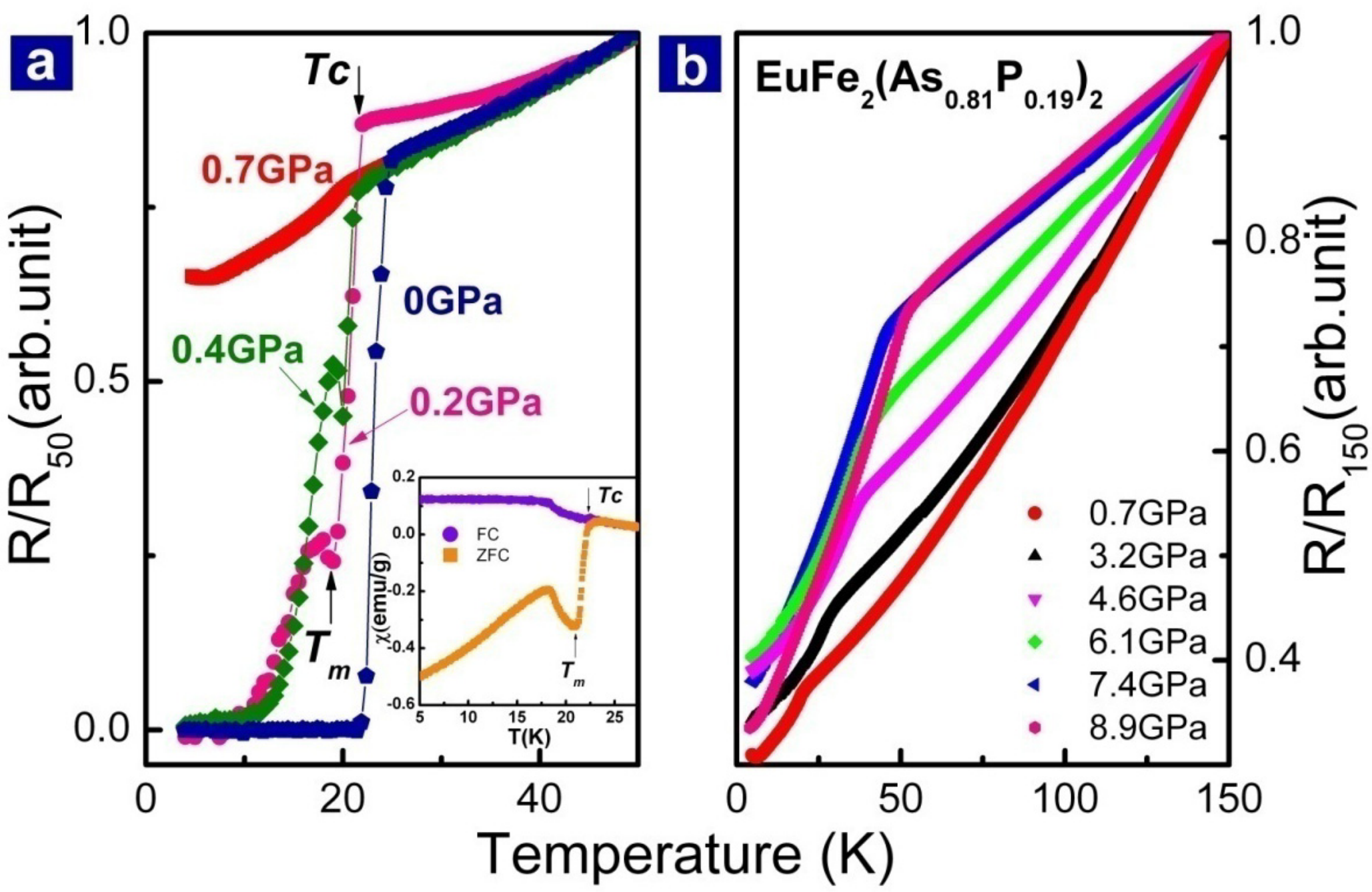

FIG.1 (Color online) Electrical resistance as a function of temperature for $EuFe_2(As_{0.81}P_{0.19})_2$ single crystal at different pressures. The inset displays dc magnetic susceptibility as a function of temperature, demonstrating the coexistence of superconductivity and a magnetic order.

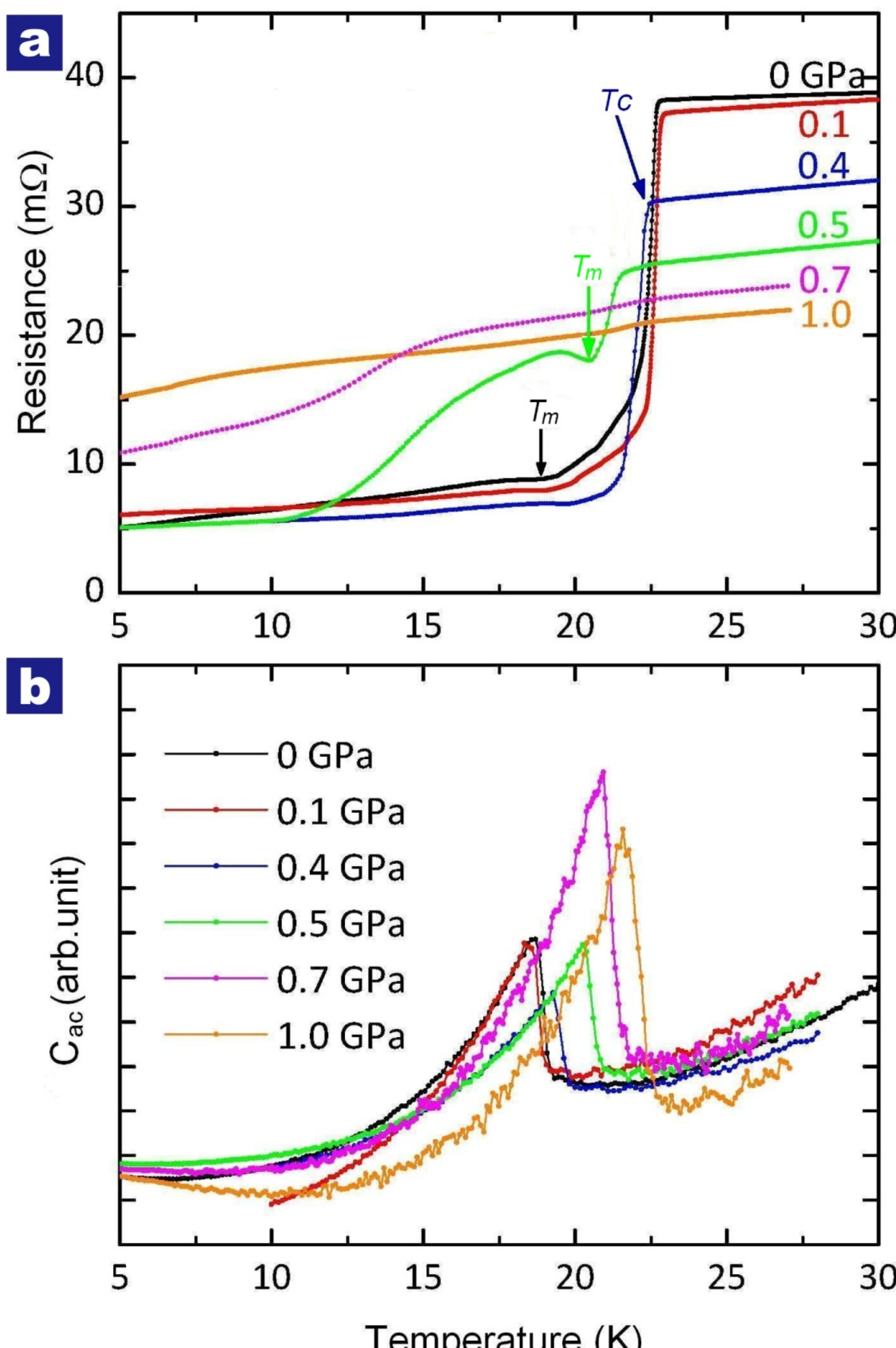

FIG.2 (Color online) Temperature dependence of (a) electrical resistance and (b) *ac*-calorimetric measurements of $EuFe_2(As_{0.81}P_{0.19})_2$ single crystal under pressure below 1GPa.

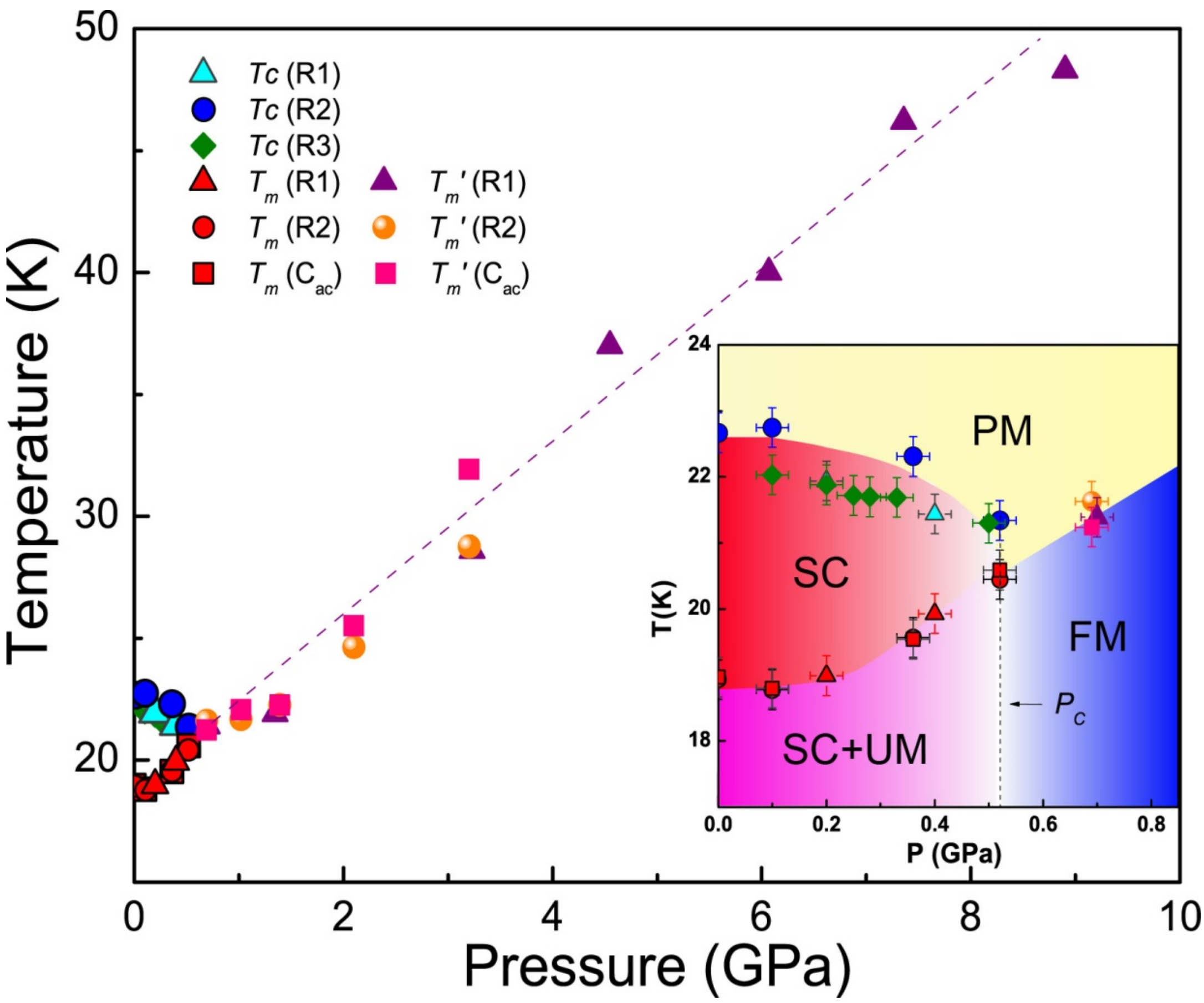

FIG.3 (Color online) Pressure-Temperature phase diagram of $EuFe_2(As_{0.81}P_{0.19})_2$ sample. The *Tc(R1)*, *Tc(R2)* and *Tc(R3)* stand for superconductivity transition temperatures obtained from different runs' resistance measurements, $T_m$ *(R1)*, $T_m$ *(R2)* and $T_m$ *($C_{ac}$)* represent onset temperature of PM-UM transitions of the intercalated layers, which are determined from different runs' resistance and *ac*-calorimetric measurements, respectively. $T_m'$ *(R1)*, $T_m'$ *(R2)* and $T_m'$ *($C_{ac}$)* represent onset temperature of PM-FM transitions, which are determined from different runs' resistance and *ac*-calorimetric measurements, respectively. *Tc (R2)* are obtained from the sample subjected to hydrostatic pressure, and *Tc(R1)* and *Tc(R3)* are obtained from the sample subjected to quasi-hydrostatic pressure.

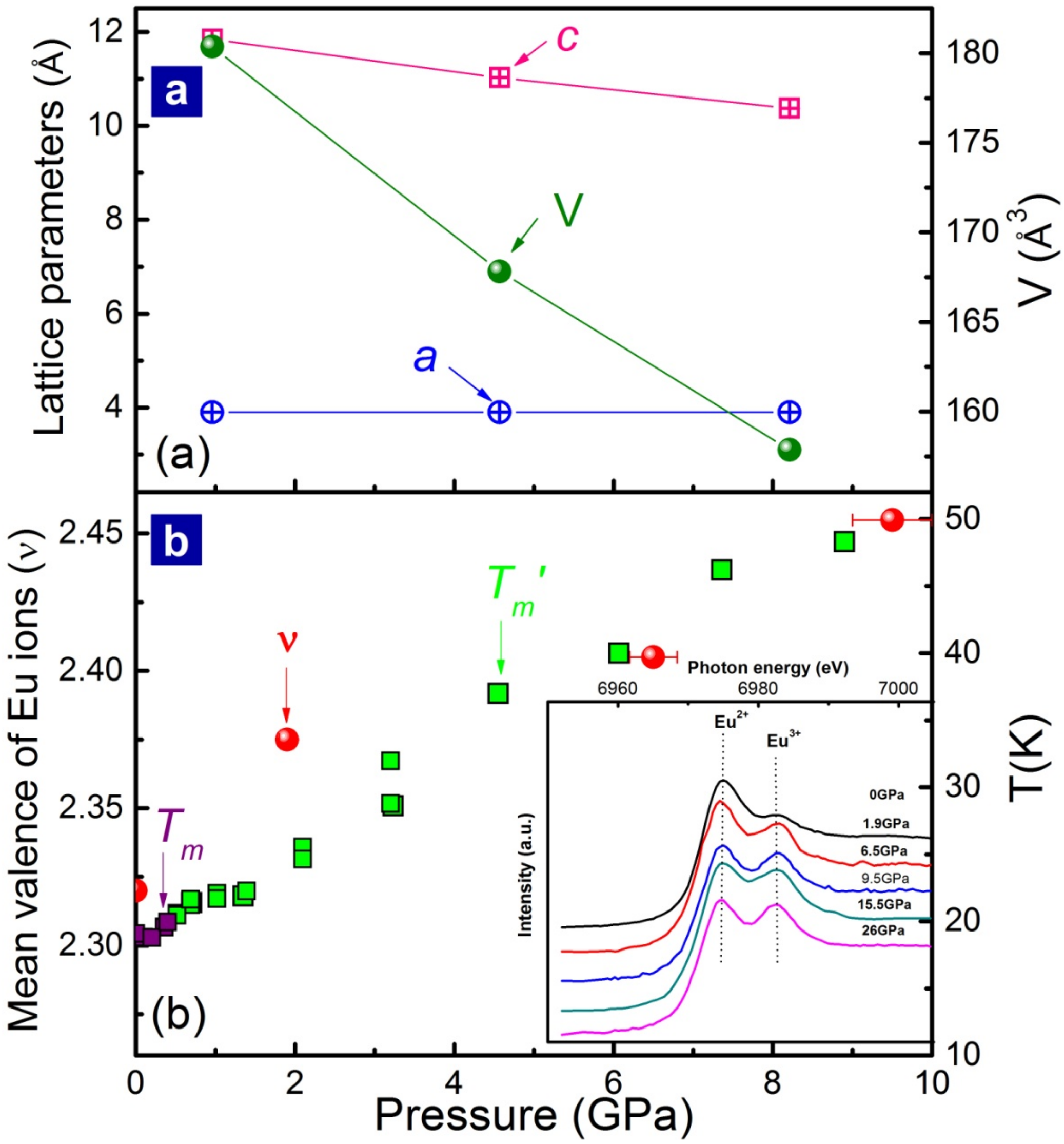

FIG.4 (Color online) (a) Pressure dependence of lattice parameters (*a* and *c*), and unit cell volume (V). (b) Pressure dependences of mean valence (ν) of Eu ions, onset temperature ($T_m$) of PM-UM transition and onset temperature ($T_m'$) of PM-FM transition temperature. The inset displays X-ray absorption spectra obtained at different pressures.